THE EUROPEAN
PHYSICAL JOURNAL C

Regular Article - Theoretical Physics

# Dealing with data gaps for TianQin with massive black hole binary signal

Lu Wang, Hong-Yu Chen, Xiangyu Lyu, En-Kun Li[a], Yi-Ming Hu[b]

MOE Key Laboratory of TianQin Mission, TianQin Research Center for Gravitational Physics and School of Physics and Astronomy, Frontiers Science Center for TianQin, Gravitational Wave Research Center of CNSA, Sun Yat-sen University (Zhuhai Campus), Zhuhai 519082, China

**Abstract** Space-borne gravitational wave detectors like TianQin might encounter data gaps due to factors like micrometeoroid collisions or hardware failures. Such events will cause discontinuity in the data, presenting challenges to the data analysis for TianQin, especially for massive black hole binary mergers. Since the signal-to-noise ratio (SNR) accumulates in a non-linear way, a gap near the merger could lead to a significant loss of SNR. It could introduce bias in the estimate of noise properties, and the results of the parameter estimation. In this work, using simulated TianQin data with injected a massive black hole binary merger, we study the window function method, and for the first time, the inpainting method to cope with the data gap, and an iterative estimate scheme is designed to properly estimate the noise spectrum. We find that both methods can properly estimate noise and signal parameters. The easy-to-implement window function method can already perform well, except that it will sacrifice some SNR due to the adoption of the window. The inpainting method is slower, but it can minimize the impact of the data gap.

## 1 Introduction

Space-borne gravitational wave (GW) detectors like TianQin [1] and LISA [2] have the potential to greatly expand and even revolutionize our observational capabilities to our Universe, enabling the detection of GW signals in the millihertz (mHz) frequency band. Detecting and analyzing these signals will illuminate the fundamental properties of the Universe and provide distinctive insights into the nature of spacetime.

[a] e-mail: lienk@sysu.edu.cn (corresponding author)
[b] e-mail: huyiming@sysu.edu.cn (corresponding author)

TianQin is a space-borne GW mission comprising three satellites. The mission is expected to be launched in the 2030 s. These satellites will form an equilateral triangle constellation in a geocentric orbit, with a radius of about 100, 000 km and an arm-length of roughly 170, 000 km. The constellation's plane will be oriented towards the reference source RX J0806.3+1527 binary system [1]. The mission will operate on a "3 months on + 3 months off" scheme [1] to avoid thermal load on the optical system. Its goal is to detect GW sources within the frequency band of 0.1 mHz to 1 Hz, including the coalescence of massive black hole binaries (MBHBs) [3–6], the inspiral of double white dwarfs [7,8], stellar-mass binary black hole inspiral [9,10], extreme mass ratio inspirals (EMRIs) [11], and the stochastic GW background [12,13].

Among these sources, MBHB is one of the primary objectives for space-borne detectors. Observational evidence indicates the ubiquitous presence of massive black hole (MBH) in the nuclei of galaxies, and it is hypothesized that triggered by the galaxy collisions, these central MBHs will ultimately be drawn together and collide [14,15]. However, our understanding of the formation and growth of the MBHs are constantly challenged and refreshed by the observations (e.g. [16]). Detecting gravitational waves from MBHBs has the potential to pinpoint the formation mechanism [3,17], depict properties of black holes in details [18,19], and decipher the growth history of the Universe [20].

The aforementioned scientific promises are often predicated upon idealized assumptions on the GW data, like the Gaussianity and stationarity. However, GW detectors encounter challenges due to periods of missing information during observations, commonly referred to as "data gaps" and thus the data do not meet the prerequisites of Gaussianity and stationarity any longer [21]. The data gaps could be raised from scheduled and unscheduled factors. Scheduled gaps are typically due to device maintenance work. For instance, laser interferometer gravitational-wave observatory (LIGO) may suspend observations temporarily for instrument upgrades and repairs, resulting in scheduled gaps [22]. Similarly, LISA may experience scheduled gaps during high-gain antenna repointing and laser frequency relocking processes [23,24], as well as periodic test mass (TM) discharge processes to get rid of charges accumulated through the interaction with the cosmic rays [25–28]. Taiji mission, similar to LISA, which also follows a heliocentric orbit but ahead of the Earth by about 18°–20°, exhibits the same gap condition as LISA [29,30]. Unlike the LISA mission, TianQin operates in a geocentric orbit, with its antenna consistently pointing toward the Earth, eliminating the need for periodic antenna adjustment [31]. Furthermore, a continuous discharge strategy could be adopted to address the issue of charge accumulation on the TMs [32]. Therefore, TianQin could avoid scheduled data gaps during observations in the ideal scenario. Unscheduled gaps may result from equipment failure or external disturbances, such as seismic interference [33] and micrometeorite collisions [34–37]. In addition, non-stationarity of the data could appear in the form of glitches, which have been widely observed in GW detectors' data [38–42]. A straightforward way to deal with glitches is to gate the data, effectively introducing gaps in the data.

Until now, most of the GW events reported in the ground-based detectors are much shorter than a typical gap. It is reasonable to assume that a signal is either totally missed in the gap or fully present in the continuous data stream. However, for space-borne GW missions, most of the signals could persist in the data for extended periods, ranging from several hours to even years. This extended duration significantly increases the probability of data gaps intersecting with these signals, thereby posing a greater challenge in terms of data continuity and signal analysis [21,43]. For example, the data gap could lead to spectral leakage in the frequency domain, contaminating other signals [44]. Secondly, the estimation of the noise's power spectral density (PSD) might need constant updates, due to the changing antenna pattern relative to the anisotropic Galactic foreground, which renders the time-dependant PSD. If the gap frequently happens, many of the data chunks would not be long enough to cover the lower end of the GW spectrum, making the task of PSD estimation much harder. The issue becomes more challenging for MBHBs, which accumulates signal-to-noise ratio (SNR)in a highly non-linear fashion as the merger time approaches [4]. A data gap occurring during the merger would result in a significant SNR loss, potentially leading to significant bias in parameter estimation [21].

There have been several pioneering works that aim to alleviate the effect of data gaps for space-borne GW detectors. This method is simple to operate, however, the suppression of spectral leakage is not complete and may result in the loss of data [45] and disruption of the Gaussian stationary characteristics of the noise. The window function method was the earliest to be adopted and simplest to implement [21,23,44,46], the major drawback is it will result in the loss of valid data points. Earlier studies like Baghi et al. [44], Carré and Porter [23] claim that using the window function method for gap data processing introduces a bias that can be as large as 9%. However, such a claim was not universally accepted, other studies like Dey et al. [21] indicate that the window function method can be bias-free. The Bayesian data augmentation (DA) method [43,47] can update the guess of the missing data, which allows the estimate of the noise PSD and GW signals simultaneously. A third method is based on the sparsity framework [48], which allows estimation of both the signal and noise, followed by gap filling with the estimated data. Both the DA method and sparsity framework method sample the signal and the missing data, simultaneously. By doing so one can naturally account for the impact of the missing data through marginalization, with the cost of extra computational burden.

While these methods provide valuable solutions for managing gapped data in space-borne GW detectors, they come with challenges. Particularly, a trade-off exists between computational complexity and accuracy. Less computationally complex methods often struggle to ensure accurate likelihood calculation.

In recent years, the gating-and-inpainting technique has been applied to handle missing data issues for ground-based GW detectors, either for the treatment of noise transients [45,49–51], or for delicate analysis of the ringdown signal [52–54]. Typically, it involves a "gating" process to eliminate unwanted data segments [55], followed by the "inpainting" stage for data interpolation. Mathematically, the inpainting technique that interpolates data and fills the gap is equivalent to masking the covariance matrix in likelihood calculation, so that the interpolated data does not bias the likelihood of the whole data. In principle, this method retains all information within the data and introduces minimal bias in the parameter estimation. The method naturally retains the continuity of the data, and it has the additional advantage of retaining all existing useful data.

In this work, we aim to investigate whether the easy-to-implement window function methods can be bias-free. We also adopt the gating-and-inpainting method to deal with the gap in space-borne GW detector data for the first time. Since the gap *per se* already acts as a gating process, hereafter we refer to the method as the "inpainting" method.

The paper is organized as follows. In Sect. 2, we introduce the theoretical background of the study. include the data stream for TianQin and the theory of parameter estimation. In Sect. 3, we detail the method and present both the window function and the inpainting technique to solve the data gap problem. In addition, the iterative algorithm estimating the noise PSD is included. In Sect. 4, We present the results of several case studies with different gap conditions for an MBHB merger. Our conclusion and discussion are made in Sect. 5.

## 2 Theoretical background

### 2.1 Data stream for TianQin

In the present paper the IMRPhenomD waveform [56,57] is used to generate the injected MBHB signals. The GW signal of a MBHB merger can be described by the following parameters: the luminosity distance $D_L$, the redshifted chirp mass $M_c$, the symmetric mass ratio $\eta$, the merger time $t_c$, the inclination angle $\iota$, the polarization angle $\psi$, the ecliptic longitude $\lambda$, the ecliptic latitude $\beta$, and the coalescence phase $\phi_c$.

The complete data **d** recorded by a GW detector is sampled at regular intervals of $\Delta t$ throughout data collection time $T$, where $T = (N-1)\Delta t$. And it can be represented as $\mathbf{d} = \{d_1, d_2, \ldots, d_N\}$. GW data consists of both signal and noise,

$$\mathbf{d}(t) = \mathbf{h}(t, \theta) + \mathbf{n}(t), \tag{1}$$

where $\mathbf{h}(t, \theta)$ represents the signal that depends on a set of parameters $\theta$, $\mathbf{n}(t)$ is noise.

The response channels of detectors we adopted are time-delay interferometry (TDI) channels: A, E, and T [10,58], which could be constructed by three Michelson-like variables, the X, Y, Z channel:

$$A = \frac{Z - X}{\sqrt{2}}, \quad E = \frac{X - 2Y + Z}{\sqrt{6}}, \quad T = \frac{X + Y + Z}{\sqrt{3}}. \tag{2}$$

The Michelson $X$ channel [59] is composed of:

$$\begin{aligned}\mathbf{X}(t) = &[\Phi_{12}(t - 3L/c) + \Phi_{21}(t - 2L/c) \\ &+ \Phi_{13}(t - L/c) + \Phi_{31}(t)] \\ &- [\Phi_{13}(t - 3L/c) + \Phi_{31}(t - 2L/c) \\ &+ \Phi_{12}(t - L/c) + \Phi_{21}(t)],\end{aligned} \tag{3}$$

here $\Phi_{ij}$ represents the recorded laser phase from satellite $i$ to satellite $j$, and $L$ is the arm-length of TianQin. The other Michelson observables $Y$, $Z$ can be obtained through cyclic permutation.

The noise model adopts TDI channel (A, E) expressed in relative frequency deviation [60]:

$$\begin{aligned}S_A(f) = S_E(f) &= 8\sin^2 u [4(1 + \cos u + \cos^2 u) \cdot S_{\mathrm{acc}} \\ &\quad + S_{\mathrm{oms}}(\cos u + 2)], \\ \sqrt{S_{\mathrm{oms}}} &= \sqrt{S_p}\frac{2\pi f}{c}, \\ \sqrt{S_{\mathrm{acc}}} &= \frac{\sqrt{S_a}}{2\pi f c}\sqrt{1 + \left(\frac{0.1\mathrm{mHz}}{f}\right)}.\end{aligned} \tag{4}$$

where $u = 2\pi f L/c$, the acceleration noise $S_a = 1 \times 10^{-30}$ m$^2$/s$^4$/Hz and position noise $S_p = 1 \times 10^{-24}$ m$^2$/Hz for TianQin. The T-channel is signal-insensitive under the low-frequency approximation and is therefore referred to as the null channel, where the signal can be ignored.

However, gapped data can be described by the product of the rectangular window function and the time-domain complete data of the TDI channel

$$\mathbf{d}_{\mathrm{gap}} = \mathbf{d} \cdot w(t), \tag{5}$$

where $w(t)$ is the rectangular window function:

$$w(t) = \begin{cases} 0, & t_{\mathrm{gap}}^i < t < t_{\mathrm{gap}}^f, \\ 1, & t < t_{\mathrm{gap}}^i, t > t_{\mathrm{gap}}^f, \end{cases} \tag{6}$$

here $t_{\mathrm{gap}}^i$ and $t_{\mathrm{gap}}^f$ represent the beginning and end of the gap respectively.

### 2.2 Parameter estimation method

Under the frame of the Bayesian inference, the posterior probability $p(\theta|\mathbf{d})$ can be expressed as:

$$p(\theta|\mathbf{d}) = \frac{p(\mathbf{d}|\theta)p(\theta)}{p(\mathbf{d})} \tag{7}$$

where $p(\mathbf{d}|\theta)$ is the likelihood, $p(\theta)$ is the prior of the parameter $\theta$, and $p(\mathbf{d})$ is the normalizing evidence.

For a Gaussian noise characterized by a $N \times N$ covariance matrix $\boldsymbol{\Sigma}$, the time domain likelihood in logarithm form can be written as:

$$\begin{aligned}\log p(\mathbf{d}|\theta) = -\frac{1}{2}\Big[&N\log 2\pi + \log|\boldsymbol{\Sigma}| \\ &+ (\mathbf{d} - \mathbf{h}(\theta))^T \boldsymbol{\Sigma}^{-1} (\mathbf{d} - \mathbf{h}(\theta))\Big].\end{aligned} \tag{8}$$

Typically, the computational complexity of inverting an $N \times N$ matrix is proportional to $O(N^3)$. This results in a large computational demand in computation time for large $N$. If the noise possesses not only Gaussian characteristics but also stationary characteristics, the covariance matrix in the time domain exhibits a Toeplitz structure, in other words, the elements along the diagonal direction are identical. Additionally, if we adopt the Whittle approximation that further treats the covariance matrix as circulant, which means that each row of the matrix is a right cyclic shift of the row above it [61], we can approximately obtain the diagonalization: $\boldsymbol{\Sigma} \approx W^\dagger \boldsymbol{\Lambda} W$, where $W$ is the discrete Fourier matrix defined as $W_{jk} = N^{-1}\exp\left(-\frac{2\pi\mathrm{i}jk}{N}\right)$, and $W^\dagger$ is the Hermitian conjugate of $W$. $\boldsymbol{\Lambda}$ possesses the nice property of being a diagonal matrix, with its eigenvalues corresponding to the noise PSD $S_n$ of the frequency domain, while the one-sided PSD $S_n(f)$ can be defined as

$$E[\widetilde{\mathbf{n}}(f_i)\widetilde{\mathbf{n}}^*(f_j)] = \frac{1}{2}TS_n(f_i)\delta_{ij}. \tag{9}$$

The inner product is defined by

$$(h_1|h_2) = 4 \sum_{i=1}^{N} \frac{h_1^*(f_i) h_2(f_i)}{S_n(f_i)} \quad (10)$$

Thus the log-likelihood can then be simplified as,

$$\log p(\mathbf{d}|\theta) \approx -\frac{1}{2}\left[ N \log 2\pi + \log |\mathbf{\Lambda}| + (\mathbf{d} - \mathbf{h}_\theta | \mathbf{d} - \mathbf{h}_\theta) \right], \quad (11)$$

By employing the frequency domain likelihood and utilizing the fast Fourier transform algorithm, the computational complexity of the covariance matrix can be significantly reduced from $O(N^3)$ to $O(N \log N)$ [62].

When the data points are missing within the time range $t \in [t_{\text{gap}}^i, t_{\text{gap}}^f]$, the affected data is presented as: $\mathbf{d}_{\text{gap}} = [d_1, \ldots, d_{t_{\text{gap}}^i}, 0, \ldots, 0, d_{t_{\text{gap}}^f}, \ldots, d_N]$. This presence of zeros disrupts the Toeplitz structure of the covariance matrix $\mathbf{\Sigma}_{\text{gap}}$. In practice, the impact of data gaps on data processing is mainly reflected in two aspects. On the one hand, data gaps can lead to spectral leakage, as discussed in Refs. [21,44,48,63]. On the other hand, due to the data gaps, the noise is no longer Gaussian and stationary, and the frequency approximation likelihood Eq. (11) is not applicable, thus necessitating the use of the time-domain likelihood function Eq. (8). Thus, the computational complexity is raised to $O(N^3)$.

## 3 Method

### 3.1 Gapped data processing

In this part, we will introduce two methods for handling the gapped data: the window function and the inpainting technique. For comparison, we also present results on data that no treatment is applied. In this case, the likelihood is evaluated with frequency-domain waveforms, with the frequency components corresponding to the gap truncated.

#### 3.1.1 Window function method

Similar to previous studies like [43], we use a Tukey-like window to handle the gapped data, as illustrated by Eq. 12.

$$w_{t_s}(t) \equiv \begin{cases} \frac{1}{2}\left(1 - \cos\left(\frac{2\pi t}{2 t_w}\right)\right) & 0 \leq t < t_w \\ 1 & t_w \leq t < t_s - t_w \\ \frac{1}{2}\left(1 - \cos\left(2\pi \frac{t - t_s + 2 t_w}{2 t_w}\right)\right) & t_s - t_w \leq t < t_s \\ 0 & \text{otherwise,} \end{cases} \quad (12)$$

$t_s$ represents the duration of each segment, and $t_w$ is the smoothing time. Mathematically, the complete window function is given by: $w_T(t) = \sum_{s=1}^{N} w_{t_s}(t - t_s)$, where $N$ represents the number of continuous data segments divided by the gaps and the function is applied to the data with gaps:

$$\mathbf{d}_{\text{win}} = \mathbf{d}_{\text{gap}} \cdot w_T(t). \quad (13)$$

The Tukey-like window function method can effectively prevent spectral leakage in the Fourier transform. It achieves this by smoothing the data at the edges of the gaps. However, this can also result in the loss of valid data points around the gaps, resulting in a compromise in SNR. The data after using the window function method can be expressed as $\mathbf{d}_{\text{win}} = [d_1 \cdot w_{t_s}, \ldots, d_{t_{\text{gap}}^i} \cdot w_{t_s}, 0, \ldots, 0, d_{t_{\text{gap}}^f} \cdot w_{t_s}, \ldots, d_N \cdot w_{t_s}]$, the covariance matrix $\mathbf{\Sigma}_{\text{win}}$ of $\mathbf{d}_{\text{win}}$ fails to satisfy the circulant property. In principle, since Whittle's approximation fails, one should carry out calculations in the time domain. However, for convenience, we still adopt Whittle's approximation and evaluate likelihood in the frequency domain.

#### 3.1.2 Gate-and-inpainting

The gate-and-inpainting technique, initially proposed by Zackay et al. [49], has been shown to effectively preserve inner product values when applied to open LIGO data [49]. This method masks the part of the covariance matrix associated with the problematic data, rendering it ineffective for matched filtering computations [51]. Isi and Farr [52] also applied the inpainting method to compute the inverse of the truncated covariance matrix. Kwok et al. [53] employed the inpainting filter to address false violations of General Relativity (GR) caused by short-duration, broadband glitches. Furthermore, Huber and Davis [50] also applied the inpainting technique to effectively mitigate the glitch problem by gating the glitch data and interpolating, reweighting, and normalizing the SNR of the gated glitch data segments. Similarly, drawing conclusions related to the quasi-normal modes requires removing portions of the dataset where the ringdown prescription is invalid, thus Capano et al. [54] applied the inpainting technique to ensure conclusions are not affected by gated inspiral-merger data. This method interpolates the data so that the likelihood after interpolation is equivalent to those before interpolation, without adding any additional information to the data.

In this work, we adopt the PyCBC realization of the inpainting to address the data gap issue Capano et al. [54]. In seeking to preserve the likelihood, the operation of data imputation from the inpainting technique helps suppress spectral leakage in the Fourier transform.

For instance, a data gap occurring during the time $t \in [t_{\text{gap}}^i, t_{\text{gap}}^f]$ in the event, the truncated noise can be represented as $\mathbf{n}_{\text{tr}} = [n_1, \ldots, n_{t_{\text{gap}}^i}, n_{t_{\text{gap}}^f}, \ldots, n_N]$. Subsequently, the log-likelihood defined in the time domain by Eq. (8) can

be expressed with:

$$\mathbf{n}_{\mathrm{tr}}^T \mathbf{\Sigma}_{\mathrm{tr}}^{-1} \mathbf{n}_{\mathrm{tr}} = \mathbf{n}'^T \mathbf{\Sigma}^{-1} \mathbf{n}', \tag{14}$$

$\mathbf{n}'$ is the noise after inpainting, $\mathbf{\Sigma}$ and $\mathbf{\Sigma}_{\mathrm{tr}}$ represent the covariance matrix of the full data and the truncated data.

To keep the value of likelihood unchanged after inpainting, it is required that

$$\begin{aligned}\mathbf{\Sigma}^{-1}\mathbf{n}'[t] &= 0, \quad t \in [t_{\mathrm{gap}}^i, t_{\mathrm{gap}}^f], \\ \mathbf{\Sigma}^{-1}(\mathbf{d}' - \mathbf{h}')[t] &= 0, \quad t \in [t_{\mathrm{gap}}^i, t_{\mathrm{gap}}^f], \end{aligned} \tag{15}$$

here $\mathbf{d}'$ and $\mathbf{h}'$ is the data and signal template after inpainting,

$$\begin{aligned}\mathbf{d}' &= \mathbf{d}_g \oplus \mathbf{d}_x, \\ \mathbf{h}' &= \mathbf{h}_g \oplus \mathbf{h}_x, \end{aligned} \tag{16}$$

where $\mathbf{d}_g$ and $\mathbf{h}_g$ are $1 \times N$ vectors, and $\mathbf{d}_x$ and $\mathbf{h}_x$ represent the data that fills in the gap with $1 \times M$ vectors. The symbols $\oplus$ indicate the filling in the vector. According to Eq. (15), $\mathbf{d}_x$ and $\mathbf{h}_x$ can be solved by:

$$\begin{aligned}\overline{\mathbf{\Sigma}^{-1}}\overline{\mathbf{d}_x} &= -\overline{\mathbf{\Sigma}^{-1}\mathbf{d}_g}, \\ \overline{\mathbf{\Sigma}^{-1}}\overline{\mathbf{h}_x} &= -\overline{\mathbf{\Sigma}^{-1}\mathbf{h}_g}. \end{aligned} \tag{17}$$

The overbar indicates the $[t_{\mathrm{gap}}^i, t_{\mathrm{gap}}^f]$ rows (and columns) of the given vector (matrix). The advantage of using inpainting is that it involves solving for a $M \times M$ matrix rather than a $(N - M) \times (N - M)$ matrix. Since the number of gapped data $M$ is much smaller than the total number $N$, this greatly reduces the computational burden. We used the PyCBC software to address inpainting data [54]. When solving for the inverse of the covariance matrix, the Levinson recursion was used to tackle the linear equation system. As a result, the computational complexity of the inpainting technique is reduced to $O(M^2)$ instead of $O(M^3)$.

### 3.2 Iterative PSD estimation

To obtain the unbiased estimation of the physical parameters of the sources, especially when the verdict of the gravity theory is at stake, accurate estimation of the noise PSD becomes a crucial task. Even though experimentalists will provide detailed modeling of the PSD, the actual onboard performance of the instruments can hardly be modeled reliably. Therefore, it is essential to estimate it from the actual data. Currently, noise PSD estimation can be categorized into two methods: off-source and on-source [13,64–68].

The "off-source" data often employs the Welch method [69] to estimate the PSD from data around, but without, the source. This method averages the periodograms of several data segments, each of the same length as the signal while excluding the signal itself [70]. Under the assumption that the off-source data contains no GW signals, and the property of the PSD remains stable within a short enough time, this method provides a reliable estimate for the PSD. However, for space-borne GW detectors, the presence of overlapping long-lasting GW signals in the data makes the previous assumptions invalid, therefore one can not simply adopt the method.

The "on-source" data can be implemented in two ways: the first is the sequential estimation, where noise parameters are iteratively estimated from the data until a stable point estimation of the noise PSD is achieved. Subsequently, the posterior probability distribution of the signal parameters is explored using the point estimation of noise PSD, as has been implemented in BayesWave [23,65–67] and FastSpec [71]. The other approach is simultaneous estimation, which explores the posterior probability distributions of both the signal and noise parameters simultaneously. Comparing the two methods, sequential estimation can provide conditional probabilities for source parameters, while simultaneous estimation provides marginal probabilities. Studies performed on ground-based detectors indicate that the above two have little difference, the peak position and the width of the posterior probabilities of the two methods differ by only a few percent [72–74].

Out of simplicity, in this study, we adopt the sequential estimation method to achieve a more rapid PSD estimation from gapped data.

The algorithm splits into five steps, see Fig. 1:

**Step 1 Data generation:** Randomly generate noise according to an inject PSD, and inject MBHB merger signal, the combination is the simulated data. At this stage, an initial guess of the PSD is also needed, but it is not necessarily similar to the inject PSD.

**Step 2 Update data:** The gapped data is processed using the window function method and inpainting method.

**Step 3 Update signal:** Using an optimizer to search for the best-matched signal template under the current knowledge of the PSD.

**Step 4 Update PSD:** Update the estimate of the noise PSD by analyzing the template-removed residual, under the current knowledge of the best-matched signal.

**Step 5 Convergence judgment:** This step controls the termination condition of the loop, to check if the estimation of the signal approaches stable, alternatively consecutive "update signal" stages give very similar estimates. If so, end the loop; otherwise, repeat from Step 2.

To search for the maximum of the posterior, we adopt the Particle Swarm Optimization (PSO) algorithm [75–80]. In this algorithm, particles representing solutions iterate through the solution space, refining their search directions based on individual and collective experiences. Ultimately, this process leads to the identification of potential optimal or nearly optimal solutions. The formula for updating particle positions is shown as follows. For the $i$-th particle, its position $x_{i,j}$ and velocity $v_{i,j}$ are updated in each iteration,

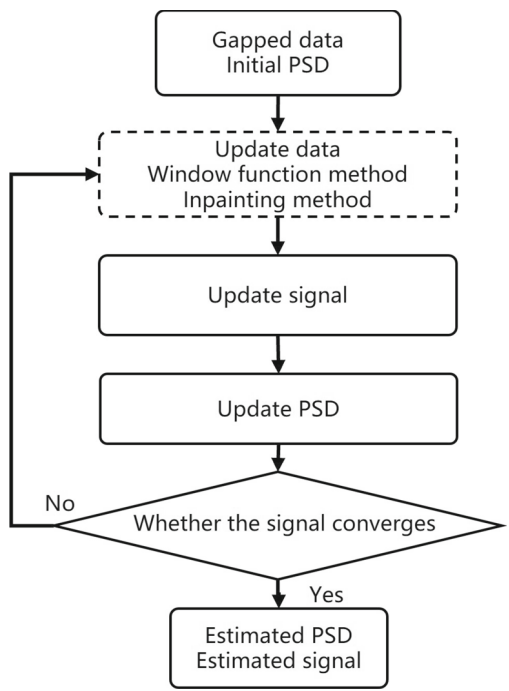

**Fig. 1** Workflow of the iterative algorithm

where the subscript $j$ denotes the component of each vector:

$$v_{i,j}(t+1) = w \cdot v_{i,j}(t) + c_1 \cdot r_{1,j}(t) \cdot [p_{\text{best},i,j} - x_{i,j}(t)] \\ + c_2 \cdot r_{2,j}(t) \cdot [g_{\text{best},j} - x_{i,j}(t)], \quad (18)$$
$$x_{i,j}(t+1) = x_{i,j}(t) + v_{i,j}(t+1),$$

where $w$ represents the inertia weight, $c_1$ and $c_2$ are the learning factors. The velocity update equation incorporates independent random numbers $r_{1,j}$ and $r_{2,j}$ for each particle as well as each component, $p_{\text{best},i,j}$ is the individual best position of the particle, and $g_{\text{best},j}$ is the global best position of the entire group.

It is worth noting that the PSO algorithm has several variants that often exhibit superior performance in tackling complex optimization problems. Such as the local best (lbest) PSO [76], which updates the local best position by restricting each particle's neighborhood. In this study, we adopt the global best (gbest) PSO algorithm; however, future research could explore the potential of the lbest PSO variant.

Inspired by the theoretical expression of the noise PSD, we perform a parametric estimation of the PSD, specifically targeting the position and acceleration noise parameters $S_a$ and $S_p$. The estimation of these noise parameters is conducted using the PSO method, as detailed in Eq. (4). Notice that the methodology is general and can be extended to a less model-dependent style.

The termination condition of the iterative algorithm relies on the convergence of the signals in two successive loops, and the fitting factor (FF) [81] is used to test the convergence of GW signal. FF is a measure used to evaluate how similar two waveforms are. The formula for calculating the FF is as follows:

$$FF = \frac{(h_1|h_2)}{\sqrt{(h_1|h_1)(h_2|h_2)}}. \quad (19)$$

Numerically, FF is limited in the range between -1 and 1, where 1 indicates a perfect match. The higher the degree of matching, the higher the degree of similarity between the assumed signal and the observed data. Termination occurs when the FF between the two signals reaches the specified threshold, the threshold for the FF is set as 0.99 in this paper.

## 4 Estimation results

As we know, the SNR of an MBHB event will increase non-linearity as it approaches the merger phase. And most SNRs are gained about 1 h before the merger [3,5]. Thus, if the gaps arise during the early inspiral phase, they are likely to have a minimal impact on the SNR of the event. To assess the impact of the large detection gap on parameter estimation. We performed a parameter estimation analysis to evaluate the impact of a 3-month gap introduced by the "3 months on + 3 months off" scheme on the early inspiral of the MBHB.

To build on this, we performed a detailed analysis to further evaluate the effects of data gaps on the short-term merger of MBHB and to evaluate the effectiveness of different gap-handling methods. To this end, we selected a source that entered TianQin's sensitive frequency band roughly two weeks before the merger. For this analysis, we used 3 days of data, which allowed us to focus more on the region near the merger. Additionally, since each likelihood function calculation requires the time-frequency transformation of the data, using 3 days of data significantly improves the computation speed. We further analyzed the effect of a 10-min gap occurring at different positions, specifically near and far from the merger. The 10-min gap is chosen to represent typical minute-scale disruptions, such as those caused by micrometeoroid collisions, and is selected to enable comparison with previous studies [21,44,48,63]. Statistical data suggest that approximately 11 such micro-meteoroid collisions may occur during a 5-year observation period of TianQin [82], equating to a frequency of roughly one event every 6 months. Therefore, we assume that such a gap could occur once during a 3-day observation period. We also present the results obtained by no treatment, the window method, and the inpainting method, respectively.

Additionally, we evaluate how the gap duration and frequency affect parameter estimation and validate the effectiveness of the method, while also discussing the effect of the same gap scenarios on parameter estimation for different chirp-mass sources.

### 4.1 Effects of a large gap at early inspiral

The TianQin project employs a "3 months on + 3 months off" observation cycle, resulting in a 3-month gap in the 9 months of data collection. Assuming an MBHB event merges on the 80th day after the start of TianQin's observation period, the parameters of this event are detailed in Table 1. The SNR of the full signal is about 818, with continuous 9-month data.

To assess the effect of the large detection gap on parameter estimation, we have compared results from two datasets: one following the 3 + 3 schedule, and another representing an ideal scenario with continuous 9-month data collection. We adopt the emcee implementation of the affine-invariant Markov Chain Monte Carlo (MCMC) sampling algorithm [83] to stochastically sample the posterior distribution. The parameter estimation results of the 11 parameters are shown in Fig. 2.

The orange contour line represents the posterior distribution obtained from 6 months of data based on the "3 months on + 3 months off" scheme. The black contour line indicates the posterior distribution from continuous observations. Blue lines which indicate the injected physical parameters are also drawn. In the figure, corner plots were shown for the marginalized two-dimensional posterior distributions, the inner and outer contour lines represent 50% and 90% credible regions, respectively. On the top, the one-dimensional marginalized distributions were also shown. We can observe that in all 1D and 2D plots, the injected values are correctly recovered within the 50% credible region. Importantly, the parameter estimation results from both schemes are highly consistent. It can be inferred that for the early inspiral of the MBHB, the parameter estimation results are not significantly affected by a 3-month observation gap introduced by the "3 months on + 3 months off" scheme.

### 4.2 Bias and error due to gap positions

Although the long gaps induced by the observation scheme have minimal impact on the early inspiral, we performed a detailed analysis to further evaluate the effects of data gaps on the short-term merger of MBHB and to evaluate the effectiveness of different gap-handling methods. To this end, we increased the chirp mass of source 1 in Table 1 by one order of magnitude, so that it enters TianQin's sensitive frequency band approximately two weeks before the merger. The SNR of the full signal is 778. To accelerate the parameter estimation process and focus more on the region near the merger, we used 3 days of data, the merger occurs at 207,360 s. As revealed in previous studies, the result of the analysis is susceptible to the position of the gap, since the accumulation of SNR becomes much quicker as the MBHB approaches the merger. A 10-min gap near the merger represents a much larger SNR loss compared with a gap far away from the merger. Therefore, we decided to study two scenarios as illustrated in Fig. 3. In the first case, the gap happens far away from the merger (the left panel), and in the other case, the gap occurs adjacent to the merger, with the ending time coinciding with the merger time (the right panel).

#### 4.2.1 PSD estimate from gapped data

In this part, we tested the iterative algorithm for estimating the noise PSD from gapped data. We injected the signal from Fig. 3 to noise and added corresponding gaps. When the gap happens far from the merger, the SNR is 777.96, almost identical to the full SNR. When the gap happens close to the merger, the SNR becomes 487.87, much smaller than the full SNR. As indicated in the flowchart (Fig. 1), an initial PSD is needed as a starting point, and we start with an extremely biased guess of the PSD, $1/S_n(f)$.

The estimated PSDs are shown in Fig. 4. Solid lines represent the results for the gap near the merger and dashed lines represent the results for the gap far from the merger. Results from no treatment, window function method, and inpainting method are indicated by black, red, and blue lines. The true value of the PSD is shown as the green solid line. For either case, when no treatment is applied to the gap, the PSD estimation is biased, especially for the $S_a$, as the impact is most significant in the low frequency. The relative bias of $S_a$ is 0.8% and 4.3%, corresponding to the near-merger and far-merger cases, respectively. On the other hand, utilizing the window function method and the inpainting method yields very similar values for both $S_a$ and $S_p$ to the injected value. The relative errors for the window function method and inpainting method are 0.01% and 0.12%, and 0.27% and 0.09%, respectively, corresponding to the near-merger and far-merger cases. We conclude also that the algorithm is very robust against the choice of the initial PSD.

#### 4.2.2 Sky localization estimation

Once we obtain an estimation of the noise PSD, we continue to perform parameter estimation of the MBHBs. Here we present the results for gapped data obtained with no treatment, the window method, and the inpainting method, separately. In this work, to make a fair comparison with previous studies like [43,50], we only focus on the sky localization parameters. For this purpose, we first show the posterior distribution for the source sky localization in Fig. 5. The dashed (solid) lines represent a gap located distant from (adjacent to) the merger, while the red and blue lines correspond to the 1 $\sigma$ confidence intervals obtained using the window function method, and the inpainting method, respectively. Gray dashed lines which indicate the injected physical parameters are also drawn. The parameter estimation errors and 1 $\sigma$ confidence intervals from the gapped data with no treatment case

**Table 1** The parameters of the injected signal

| Parameters | Values | Parameters | Values |
| --- | --- | --- | --- |
| $M_c/M_\odot$ | $3 \times 10^5$ | $t_c$ | 80 days |
| $\eta$ | 0.15 | $D_L$/Gpc | 10 |
| $\chi_1$ | 0.5 | $\lambda$/rad | $\pi$ |
| $\chi_2$ | 0.1 | $\beta$/rad | $\pi/3$ |
| $\phi_c$/rad | $\pi/2$ | $\psi$/rad | $\pi/4$ |
|  |  | $\iota$/rad | $\pi/8$ |

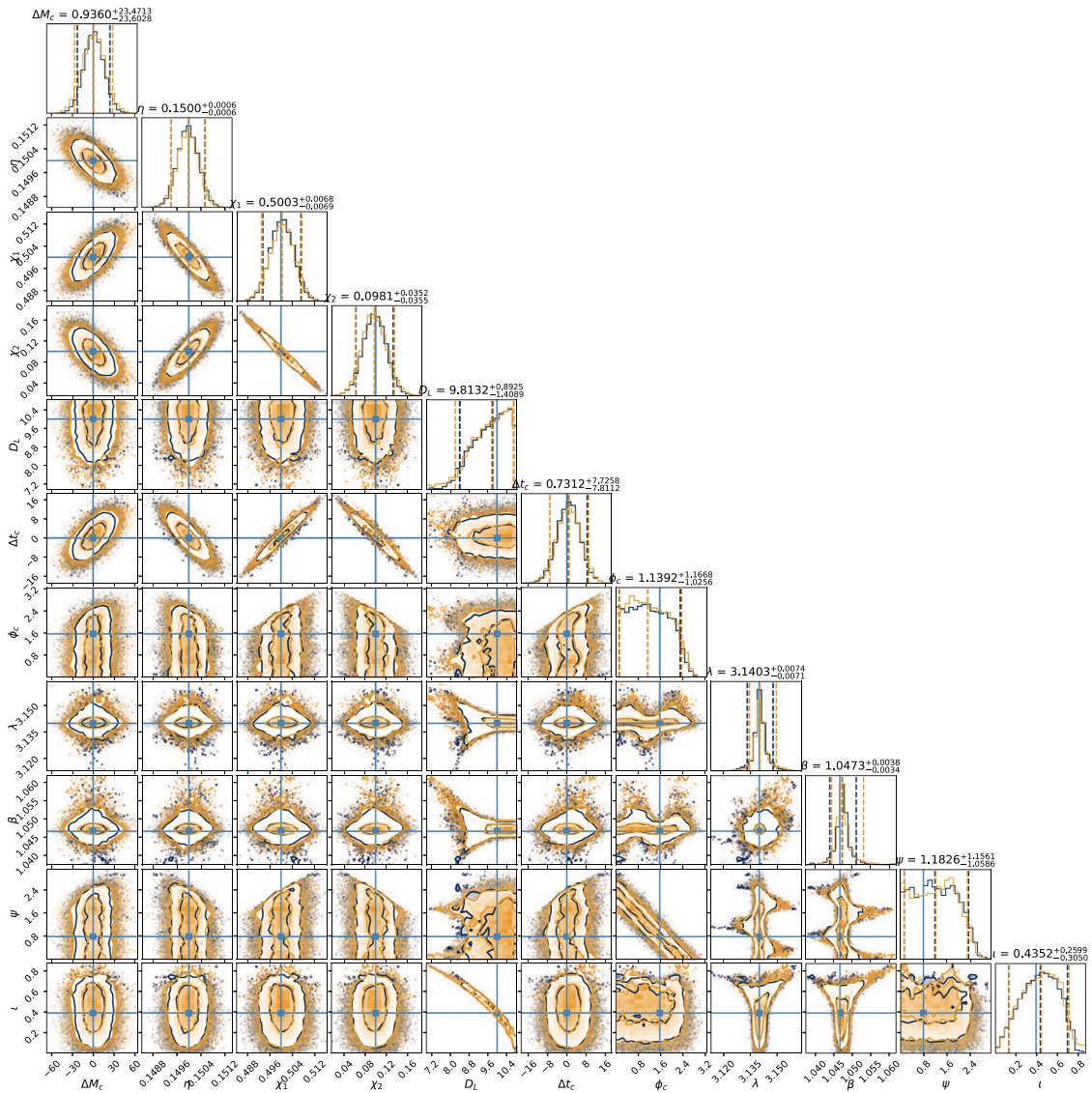

**Fig. 2** The 11-dimensional parameter estimation results for "3 months on + 3 months off" scheme and continuous observations. The orange contour line is the posterior distribution obtained from data with the "3 months on + 3 months off" scheme. The black contour line indicates the posterior distribution based on continuous observations. The blue line represents true values

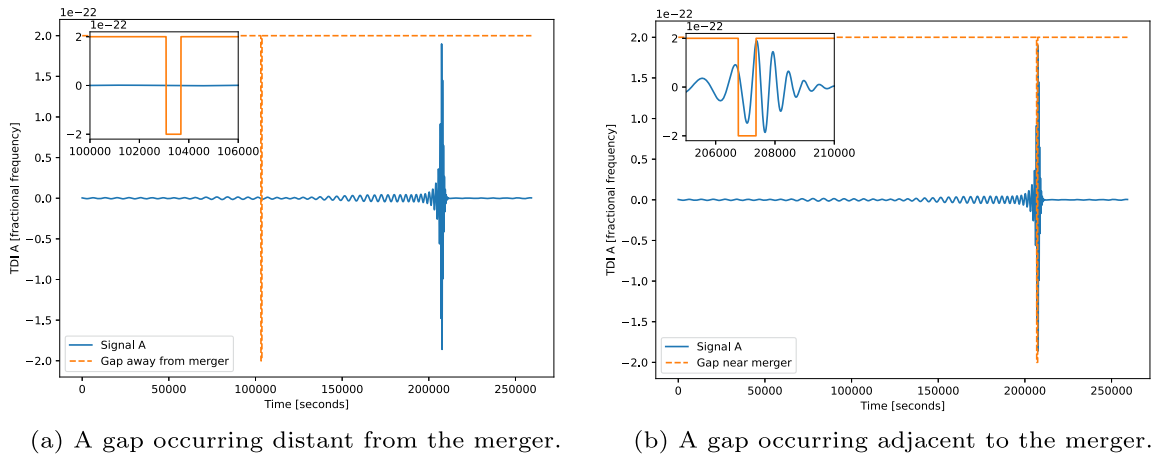

(a) A gap occurring distant from the merger.

(b) A gap occurring adjacent to the merger.

**Fig. 3** The schematic figure of time domain waveform (blue solid line) and the localization of gap (orange dashed line)

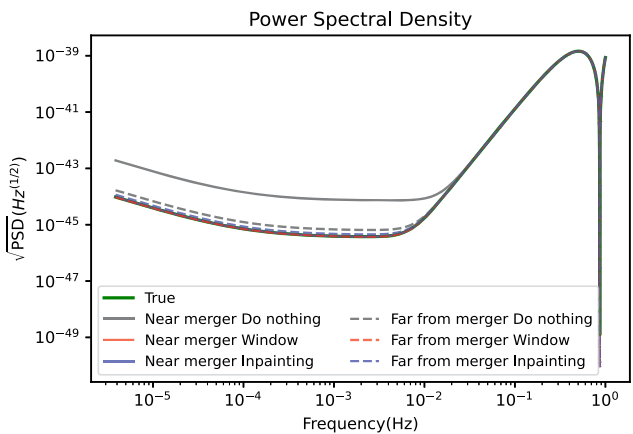

**Fig. 4** The PSD estimation results for different positions of the gap. The dashed (solid) lines represent a gap located distant from (adjacent to) the merger. The estimated PSD of no treatment, window function method, and inpainting method are indicated by black, red, and blue lines, respectively. The true value of the PSD is shown as the green solid line

differ significantly from those obtained using gap-handling methods. Therefore, they cannot be effectively presented in the same figure. Instead, the parameter estimation errors and $1\sigma$ confidence intervals for the three methods are summarized in Table 2.

We can observe that both the window function and the inpainting methods give consistent estimates, with the injected parameter constrained within the center of the credible regions. When the gap occurs far away from the merger, both the window function method and the inpainting method can effectively recover the injected parameter. The two methods result in almost identical sizes and locations for the credible regions. On the other hand, in scenarios with the gap occurring at the merger, we observe much wider credible regions for the window function method than the inpainting method. This can be explained by the fact that the SNR of the MBHB accumulates much faster around the merger. The window function methods inevitably taper some part of the data, which leads to loss of information when that data contains a significant amount of SNR, and thus the widen-

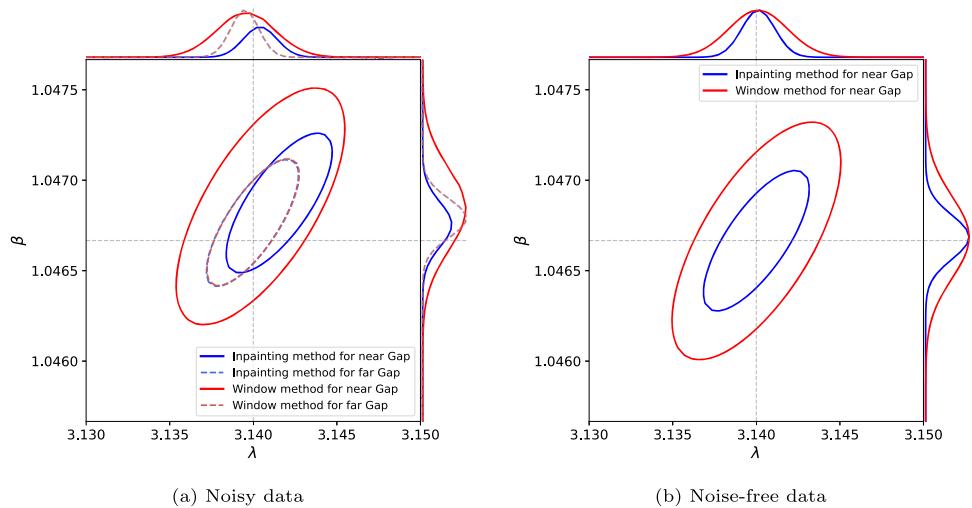

**Fig. 5** The posterior distribution for the source sky localization. The gray dashed lines represent injected values. The dashed (solid) lines represent a gap located distant from (adjacent to) the merger, and the red lines and the blue lines correspond to the $1\sigma$ confidence intervals obtained using the window function method, and the inpainting method, respectively. The left (right) of the figure represents noisy (noise-free) data

(a) Noisy data

(b) Noise-free data

**Table 2** Parameter estimation errors for gaps at different positions

| Conditions | No treatment | | Window function | | Inpainting | |
|---|---|---|---|---|---|---|
| | $\delta\lambda$ | $\delta\beta$ | $\delta\lambda$ | $\delta\beta$ | $\delta\lambda$ | $\delta\beta$ |
| Gap far from the merger | $-0.61^{+12.26}_{-12.28}$ | $0.06^{+1.52}_{-1.60}$ | $-0.20^{+2.94}_{-2.55}$ | $0.10^{+0.35}_{-0.35}$ | $-0.20^{+2.9}_{-2.59}$ | $0.10^{+0.34}_{-0.36}$ |
| Gap adjacent to the merger | $134.75^{+7.78}_{-7.22}$ | $21.85^{+1.84}_{-2.03}$ | $0.61^{+4.86}_{-5.23}$ | $0.22^{+0.62}_{-0.69}$ | $1.43^{+3.29}_{-3.04}$ | $0.18^{+0.41}_{-0.36}$ |

ing of the credible regions. The figure is obtained by setting the tapering length $t_w = 1000$ s. Finetuning the value of $t_w$ may yield quantitatively different results, but the qualitative conclusions remain unchanged.

Through numerical analysis in Table 2, the parameter estimation errors with no treatment remain relatively small even if the PSD estimation is incorrect when the gap occurs far away from the merger. However, in scenarios where a gap occurs at the merger, even if the PSD estimation is close to the injected PSD, the parameter estimation errors are up to two orders of magnitude higher than when the gap occurs far from the merger. This is because the gaps occurring at the merger location lead to more severe spectral leakage of the signal, resulting in a mismatch with the ideal frequency domain template. when the gap is far from the merger, although the errors for no treatment case and the inpainting method are similar in magnitude, the inpainting method significantly reduces the $1\sigma$ error range, decreasing it by about five times. Near the merger, for a 10-min gap, the inpainting method significantly improves the accuracy of parameter estimation, enhancing it by two orders of magnitude.

Next, we quantitatively examine the parameter estimation bias introduced by the methods. The difference between the injected parameter (gray dashed lines) and the maximum values of 1D in Fig. 5 does not overlap, and this bias can be explained by the intrinsic random fluctuation caused by the noise. However, treating the gap *per se* can potentially bring in bias. To separate this effect from the randomness raised by noise, we repeat the analysis on noise-free data and show results in Fig. 5. The maximum posterior of noise-free data for both the window function and inpainting methods are located at the same position as the true values, which indicates that the errors in both methods primarily arise from noise.

### 4.2.3 Bayesian estimation of 11 parameters with gapped data

Finally, we aim to present a more realistic and useful result, to perform parameter estimation on an 11-dimensional space. For the first time, we obtain full dimensional parameter estimation results upon a gapped space-borne GW detector data, with noise included. We show the results in Fig. 6. The data set is identical to that of Fig. 3b, with the gap happening dur-ing the merger. The gap is filled by the inpainted data constructed by the estimated noise PSD as well as the waveform corresponding to the sampled parameter set. This means that for every likelihood evaluation, the inpainting method is executed. Thanks to the high efficiency of our inpainting implementation, $5 \times 10^6$ likelihood evaluations were performed within 40 h on a 50 core cluster.

In the figure, corner plots were shown for the marginalized two-dimensional posterior distributions, the inner and outer contour lines represent 50% and 90% credible regions, respectively. On the top, the one-dimensional marginalized distributions were also shown. Blue lines which indicate the injected physical parameters are also drawn. We can observe that in all 1D plots, the injected values are correctly recovered within the 90% credible region. For the 2D plots, all injected values are recovered within the 90% credible region, while about half are within the 50% credible regions. We can compare with Fig. 5, and not surprisingly, the 2D posterior marginalized over all other dimensions is wider than the conditional 2D distribution. When comparing with previous studies on gap-free data, for example, Gao et al. [6], we can find that most of the 2D distributions are quite consistent, with most parameter precision agreeing within a factor of 3. Three parameters can differ by 1–2 orders of magnitudes: merger time $t_c$, polarization angle $\psi$, and inclination angle $\iota$. The difference in polarization angle can be explained by the fact that our observation time (3 days) is significantly shorter than [6] (3 months). The weak constraining of the polarization angle leads to a poorly constrained inclination angle as well as the luminosity distance, as the degeneracy can not be broken. For our case, the merger time can be constrained to the level of $\sim 100$ s, while in Gao et al. [6] the precision is a factor of ten smaller. This is caused by the fact that we studied data corrupted by the gap, which exactly happens during the merger, so the core information of the merger time is missing.

### 4.3 Bias and error due to different gaps

In this subsection, we evaluate how the gap duration and frequency affect parameter estimation and validate the effectiveness of the method. Additionally, we discuss the effect of the same gap scenarios on parameter estimation for different chirp-mass sources.

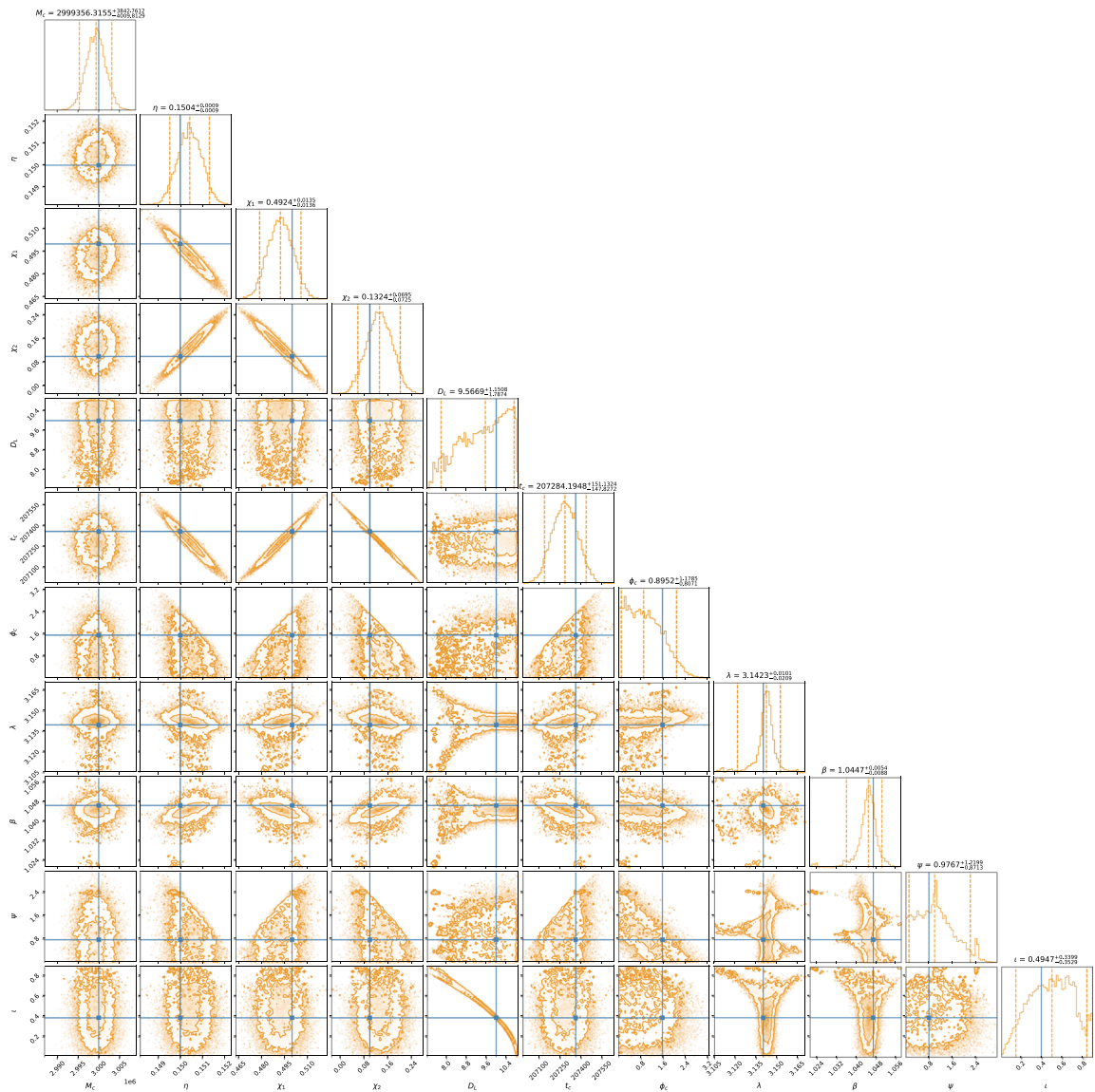

**Fig. 6** The 11-dimensional parameter estimation results using the inpainting method for the case when the gaps occur at the merger. The blue line represents true values

*4.3.1 The effect of gap duration*

We investigated the results for gaps of varying lengths. We selected three gap durations: 20 s, 5 min, and 1.5 h. These durations correspond to real-world scenarios: 20 s represent a brief glitch-induced interruption, 5 min simulate the duration required to recover to science mode using an advanced control algorithm after a micro-meteoroid collision [82], and 1.5 h represents the total duration of the interruption recovery time required for re-locking the laser, from initial scanning up to science mode, for all constellation arms [84]. Building on the results from the previous subsection, to highlight the effect of the gaps and the effectiveness of the inpainting method, we set the end times of all gaps to coincide with the merger time. We also present the results of no treatment and the inpainting method, respectively.

The PSD estimation results for the no-treatment case show varying degrees of deviation for $S_a$, with the maximum relative error reaching 2.4%. In contrast, when the inpainting method is used for PSD estimation, the deviation is much smaller, only 0.27%, improving by approximately an order of magnitude.

Next, the signal parameter estimation is performed in Fig. 7 based on the estimated PSD using the inpainting method. The light blue dashed lines correspond to the 1 $\sigma$ confidence intervals obtained with a 20-s gap, the medium blue dotted solid line represents the result for the 5-min gap, and the dark blue solid line represents the result for the 1.5-h

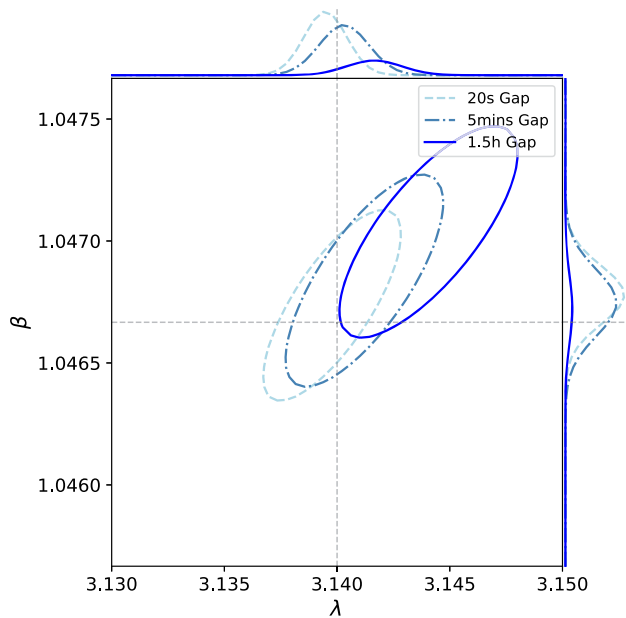

**Fig. 7** Sky localization of the data with the inpainting method. The light blue dashed lines correspond to the 20-s gap, the medium blue dotted solid line represents the 5-min gap, and the dark blue solid line represents the 1.5-h gap

gap. Table 3 presents the parameter estimation errors and 1 $\sigma$ confidence intervals for different gap lengths obtained by no treatment and the inpainting method, and is defined in the same way as Table 2.

In Fig. 7, when the inpainting method is used, the bias increases with the gap length, but the 1 $\sigma$ confidence intervals remain relatively stable, and the parameter estimates for all three gap lengths still fall within approximately the 1 $\sigma$ confidence intervals. This is because, although the inpainting method effectively mitigates spectral leakage caused by discontinuities, longer gap durations introduce greater spectral leakage, offsetting the method's effectiveness. However, the stability of the PSD estimation ensures that the error range remains small, which is why the 1 $\sigma$ confidence intervals stay relatively consistent.

As shown in Table 3, as the gap length increases, both the parameter estimation errors and the 1 $\sigma$ confidence intervals increase in no treatment case. For example, the impact of a 1.5-h gap on the parameters is nearly 8 times greater than that of a 20-s gap, with a bias of $158.08 \times 10^{-3}$. This is because the unsuppressed spectral leakage effect becomes more pronounced as the gap length increases, leading to a larger error. In extreme cases, the inpainting method reduces the error by two orders of magnitude compared to no treatment case.

### 4.3.2 The effect of gap count

In this subsection, we focus on the effect of varying gap numbers. To further emphasize the effects of the gaps, we place a single 10-min gap near the merger and consider scenarios with 1, 10, and 100 injected glitches, each lasting 20 s.

Figure 8 illustrates the estimated PSD results for different numbers of glitches using both no treatment case and the inpainting method. The lines with varying degrees of proximity to black represent the results of no treatment case, while the lines with varying degrees of proximity to blue correspond to the results obtained using the inpainting method. As the color of the lines deepens, it indicates an increase in the number of glitches, transitioning from 1 to 10 to 100. For no treatment case, the PSD estimation shows varying degrees of deviation, with the deviation increasing as the number of glitches rises. The maximum deviation reaches 8.73%, with the PSD curve deviating by four orders of magnitude at low frequencies, as shown by the dark black solid line. In contrast, the inpainting method produces a much smaller PSD estimation bias of 0.23%, offering a 37-fold improvement and closely aligning with the true PSD.

The parameter estimation results for cases with different numbers of gaps, using the inpainting method are shown in Fig. 9, where the lines transition from light green to medium green to dark green, represent the posterior distributions of the parameters for the cases injected glitch counts of 1, 10, and 100, respectively. Table 4 presents the parameter estimation errors and 1 $\sigma$ confidence intervals for different glitch counts, as defined in Table 2.

Figure 9 shows that for the inpainting method, the bias remains within the 1 $\sigma$ confidence intervals for scenarios with 1 and 10 glitches, and even in the extreme case of 100 glitches, they stay within the 3 $\sigma$ range. In contrast, for no treatment case, a single glitch results in a large bias exceeding 10 $\sigma$, while for 100 glitches, the bias unexpectedly decreases and falls within the 1 $\sigma$ confidence intervals. However, compared to the inpainting method, no treatment case shows greater deviation from the true values and a wider error range. These findings suggest that as the number of glitches increases, the gaps have a more significant effect on PSD estimation, resulting in larger deviations. The inpainting method effectively mitigates these biases compared to leaving the gaps untreated.

### 4.3.3 The effect of data gaps on different chirp masses.

To discuss the impact of gaps on different chirp mass sources, we assume that source 1 in Table 1 contains only 3 days of valid data and shares the same merger time $t_c$ as source 2, with a 10-min gap occurring near the merger, as discussed in Sect. 4.2.

**Table 3** Parameter estimation errors for different gap lengths

| Conditions | No treatment | | Inpainting method | |
|---|---|---|---|---|
| | $\delta\lambda$ | $\delta\beta$ | $\delta\lambda$ | $\delta\beta$ |
| 20 s | $18.20^{+6.07}_{-6.02}$ | $-3.13^{+0.81}_{-0.80}$ | $-0.20^{+3.03}_{-3.07}$ | $0.06^{+0.40}_{-0.38}$ |
| 5 mins | $114.49^{+4.58}_{-3.50}$ | $0.41^{+0.45}_{-0.41}$ | $1.43^{+3.28}_{-3.71}$ | $0.18^{+0.42}_{-0.45}$ |
| 1.5 h | $158.08^{+29.85}_{-29.69}$ | $23.83^{+16.16}_{-17.44}$ | $3.88^{+4.12}_{-3.77}$ | $0.35^{+0.46}_{-0.41}$ |

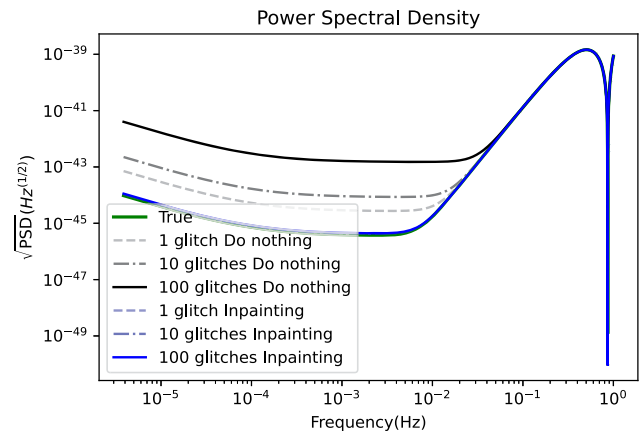

**Fig. 8** The PSD estimation results for different numbers of gaps. The lines with varying degrees of proximity to black represent the results of no treatment case, while the lines with varying degrees of proximity to blue correspond to the results obtained using the inpainting method. The light dashed line, medium-shaded dash-dot line, and dark solid line represent cases with glitch counts of 1, 10, and 100, respectively

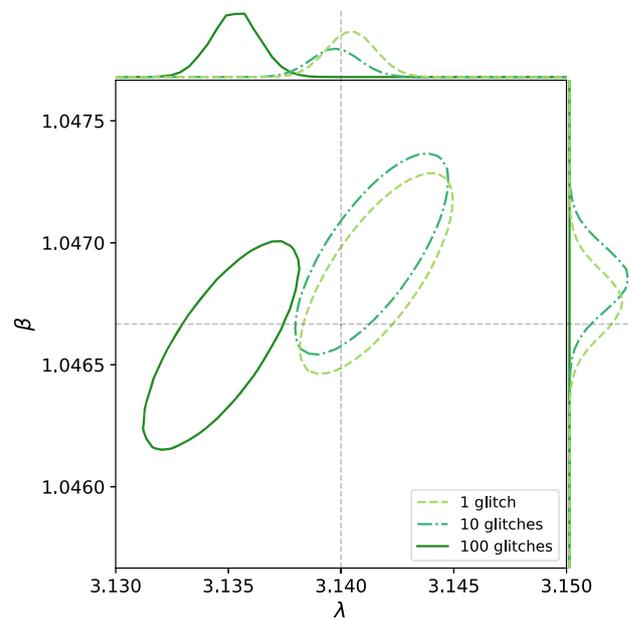

**Fig. 9** The posterior distribution for the source sky localization. The light green dashed line, the medium green dotted solid line, and the dark green solid line represent the injected glitch counts of 1, 10, and 100, respectively

The PSD estimation results align with the previous: without correction, the PSD shows a 65% deviation, while the inpainting method yields significantly smaller errors.

The parameter estimation errors and 1 $\sigma$ confidence intervals for no treatment case and the inpainting method are shown in Table 5. Compared to the results in Table 2, in no treatment case, sources with smaller chirp masses exhibit smaller biases than those with larger chirp masses for the same gap. This is because sources with relatively smaller chirp masses have a slower accumulation of SNR near the merger, resulting in a smaller SNR loss for the same gap. However, the inpainting method results in nearly consistent errors regardless of whether the mass increases or decreases, and effectively reduces the errors compared to no treatment case.

## 5 Conclusion

Space-borne GW detectors might encounter data gaps due to factors like micro-meteoroid collisions or hardware failures. Data with gaps will exhibit non-Gaussian and non-stationary characteristics, rendering theoretical noise modeling ineffective and necessitating noise estimation from the data. Additionally, gaps lead to a loss of signal information, which is especially critical during the merger phase for chirp signals like MBHBs. These signals exhibit a nonlinear SNR increase as they approach the merger phase, with most of the SNR gained about 1 h before the merger [3,5].

We performed a parameter estimation analysis using the "3 months on + 3 months off" scheme and continuous observations. The results show that the 3-month long gap induced by the observation scheme has minimal impact on the early inspiral of the MBHB. This finding suggests that it is sufficient to focus on gaps occurring near the merger phase. Therefore, our analysis emphasizes addressing the gaps in MBHB events close to the merger.

In this work, we applied the window function method, and for the first time, the inpainting method to cope with the data gap, and an iterative estimate scheme is designed to properly estimate the noise spectrum. For the PSD estimation, the no treatment case resulted in a significant PSD estimation devi-ation, whereas employing the window function and inpainting methods produced PSD estimations consistent with the injected values, with the deviation much smaller, less than 1%.

**Table 4** Parameter estimation errors for different gap counts

| Conditions | No treatment | | Inpainting method | |
|---|---|---|---|---|
| | $\delta\lambda$ | $\delta\beta$ | $\delta\lambda$ | $\delta\beta$ |
| 1 glitch | $134.24^{+13.85}_{-12.60}$ | $21.78^{+3.46}_{-3.35}$ | $1.84^{+3.12}_{-3.67}$ | $0.22^{+0.39}_{-0.43}$ |
| 10 glitches | $92.52^{+12.85}_{-13.16}$ | $1.26^{+1.41}_{-1.41}$ | $1.43^{+3.32}_{-3.46}$ | $0.31^{+0.39}_{-0.43}$ |
| 100 glitches | $50.^{+53.48}_{-53.35}$ | $-3.01^{+6.24}_{-6.67}$ | $-5.51^{+3.65}_{-3.28}$ | $0.14^{+0.37}_{-0.48}$ |

**Table 5** Parameter estimation errors for different chirp-mass

| Condition | No treatment | | Inpainting method | |
|---|---|---|---|---|
| | $\delta\lambda \times 10^{-3}$ rad | $\delta\beta \times 10^{-3}$ rad | $\delta\lambda \times 10^{-3}$ rad | $\delta\beta \times 10^{-3}$ rad |
| $3 \times 10^5 M_\odot$ | $97.76^{+10.02}_{-12.84}$ | $0.51^{+0.49}_{-0.57}$ | $1.43^{+2.15}_{-1.87}$ | $0.02^{+0.03}_{-0.04}$ |

With the PSD fixed, we perform parameter estimation for the MBHB signals. Both the window function and the inpainting method improve accuracy compared to no treatment. However, the tapering of the window function method contributes to SNR loss and consequently expands the credible range. The exact results are subject to the choice of smooth time $t_w$. As expected, when the gap occurs near the merge time of an MBHB event, the inpainting method can gain much better results compared with the window method.

Compared with the inpainting method, the window function method is easy to implement. It works well when the gap occurs far from the merger. This result differs from earlier studies but confirms observations from Dey et al. [21], while we took a step further in that realistic noise was included throughout the study. However, for gaps closer to the merger, the loss of some data points may result in an enlarged range of errors. Notably, the results of the above-mentioned windowed gapped data are based on several assumptions. In practice data processing, when employing the window function, the covariance matrix of windowed gapped data does not retain the Toeplitz property. Therefore, similar to gapped data, Whittle's approximation fails.

By employing an inpainting technique, the covariance matrix corresponding to the gaps is masked. The masked covariance matrix represents the true covariance matrix after data loss, thus preserving information on matching filter and likelihood. The computational cost of the likelihood is $O(N \log N)$, and the computational cost of inpainting is $O(M^2)$, where $M$ is the number of missing data points.

In real-world scenarios, the location, duration, and frequency of data gaps are usually unpredictable and beyond control. We investigate the results for scenarios where gaps occur at different stages of the MBHB merger, with varying lengths and numbers. Firstly, the results show that when the gap occurs far away from the merger, with a relatively small loss in SNR, the parameter estimation will still be close to the true value, even if the PSD is wrongly estimated. As the gap occurs at the merger, with a substantial loss in SNR, it leads to more severe spectral leakage of the signal. In this case, even if the true PSD is adopted, the estimated signal is still biased. For a 10-min gap near the merger, the inpainting method significantly improves the accuracy of parameter estimation, enhancing it by two orders of magnitude. Secondly, longer gaps exert a stronger influence on the data, particularly when they occur near the merger phase. Even though the inpainting method improves parameter estimation accuracy, it cannot fully eliminate errors caused by excessively long gaps. Thirdly, as the number of gaps increases, their effect on PSD estimation becomes more pronounced, resulting in greater deviations. While the inpainting method significantly reduces errors compared to cases with no treatment. In extreme cases, the inpainting method improved the estimation accuracy of parameter $S_a$ by nearly 40 times compared to no treatment case. Even in the extreme case of 100 glitches, however, the bias remains within the $3\sigma$ range for the inpainting method. Additionally, the study examined signals with different chirp masses, revealing that sources with smaller chirp masses exhibit smaller biases compared to those with larger chirp masses in the same gap case. Regardless of whether the chirp mass increases or decreases, the inpainting method consistently delivers nearly uniform accuracy, effectively reducing estimation errors compared to no treatment case.

The current limitation is that the noise model relies on a strong prior assumption, which has shortcomings in predicting or estimating the noise PSD. The estimation of noise PSD is based on maximum likelihood point estimation. Ideally, conclusions should be drawn after marginalizing over the PSD parameters, instead of based on a fixed estimate [43,72]. Unfortunately, the implementation of such a target will require significantly larger computation resources, and we leave the development towards this target to future works.

**Acknowledgements** The authors thank Yi-fan Wang and Jiandong Zhang for fruitful discussions at various stages of this work. Their expertise and valuable suggestions have provided crucial insights for this study. We also extend our gratitude to Liang Dai for equally invaluable discussions. This work has been supported by the National Key Research and Development Program of China (no. 2023YFC2206700), the Guangdong Major Project of Basic and Applied Basic Research (Grant no. 2019B030302001), the Natural Science Foundation of China (Grant no. 12173104, no. 12261131504), and the Natural Science Foundation of Guangdong Province of China (Grant no. 2022A1515011862).

**Data Availability Statement** Data will be made available on reasonable request. [Author's comment: The datasets generated during and/or analysed during the current study are available from the corresponding author on reasonable request.]

**Code Availability Statement** Code/software will be made available on reasonable request. [Author's comment: The code/software generated during and/or analysed during the current study is available from the corresponding author on reasonable request.]